\input{aipcheck}

\documentclass[
    ,final            
  ]{aipproc}

\layoutstyle{8x11double}

\begin{document}

\title{We must know. We will know.}


\author{Miguel-Angel Sanchis-Lozano}{
  address={IFIC, Centro Mixto Universidad de Valencia-CSIC, Burjassot, Spain}
}

\begin{abstract}
The after-dinner talk has by now become a tradition
of this Conference series on Quark Confinement and 
the Hadron Spectrum. 
On this occasion, I have tried to combine 
a free-style and (hopefully) amusing presentation with 
deep questions of physics especially connected
with the dynamics of strong interaction. To this end
some masterpieces of classical music
(by Beethoven, Mozart, Dvorak, Stravinsky ...) and
pop music (by Bob Dylan, Eric Clapton)
were employed to illustrate certain aspects of physics.
By no means was this presentation (neither this paper)
intended as a comprehensive
review of the different topics examined during the Conference, but 
rather as a
call for further thinking on the sinergy of different branches
of physics and the excitement of foreseen discoveries 
in a not too distant future. 
\end{abstract}

\maketitle


\begin{figure}
  \includegraphics[height=.15\textheight]{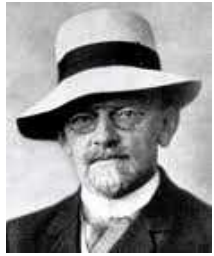}
  \caption{Picture of the great German mathematician David Hilbert. His
determination {\em to know} is a strong inspiration for this article. 
However, 
science is currently facing complex issues which concern the
essence of the scientific method itself. The dream of a 
$\lq\lq$theory
of everything'' might become a nightmare for many, when 
apparently there is no
way to find the expected $\lq\lq$true solution'' from among 
a huge set of possible vacua!}
\end{figure}

\section{Introduction}

The defiant statements entitling this paper
($\lq\lq${\em Wir müssen wissen. Wir werden wissen.}'') 
pronounced by David Hilbert 
at the 1930 Annual Meeting of the German Scientists and Physicians, 
can be viewed by many 
as too optimistic, conveying a bold belief in the
unlimited capacity of the human mind to grasp any law of Nature.
In fact, G\"{o}del had announced the first expression of his famous 
incompleteness theorem just one day before! 

In any event, one can truly marvel at the present status of 
the human understanding of Nature. With the help of sophisticated 
devices and computers for
detection and analysis (which can be interpreted as 
$\lq\lq$magnified'' senses and brain),
scientists have been able to explore 
length scales from the extremely tiny to the 
cosmologically large. This task has been accomplished throughout
history by 
a long (not yet finished), often 
hard struggle of brave people against ignorance, 
superstition and fanaticism. 

Admittedly, the advancement of science looks
now unstoppable, fueled by the technological achievements 
demanded by society. 
Nonetheless shadows still emerge from the unavoidable 
complexity of current 
scientific knowledge.   
Simpler (but false) interpretations of the natural
world (e.g. creationism, astrology, etc) may look
more appealing than conventional wisdom for people
without a scientific background.

In physics, the naive classical determinism of the 
nineteenth century has been replaced by a set of uncertainties
arising from quantum mechanics or chaotic behaviour. 
More recently, even the essence of the scientific
method has become controversial, inasmuch as the anthropic principle is 
invoked. The pending question of the vacua landscape in string theory
has indeed become a nightmare for some buoyant
believers in the existence of a (single?) $\lq\lq$theory of everything''. 
In spite of (or perhaps because of) such controversial 
situation \cite{Smolin, Penrose}, 
the future of elementary and astro-particle physics, alongside
cosmology, definitely looks exciting.

Following the order of the after-dinner talk, we briefly
examine in the next section the birth of 
systematic learning about nature in ancient Greece. 
As Aristotle put it: $\lq\lq${\em All men by nature desire to know}''.
History of science can teach us many lessons valuable today.
If we cannot be sure of where we are going, scientifically speaking, 
at least let us keep in mind from where we have come.

In the subsequent section, we contemplate the intertwining of 
the strong interaction and string theories as an example of
conceptual entanglement between different fields in physics. 
New symmetries, new possible spectroscopies corresponding to
hidden sectors beyond the Standard Model (SM) were also mentioned
during the after-dinner presentation. Finally we stress
the relationship between the $\lq\lq$large'' (astrophysical and
cosmological models)
and the $\lq\lq$small'' (nuclear and particle physics),
focusing on the detection of (cold) dark matter.

\begin{figure}
  \includegraphics[height=.23\textheight]{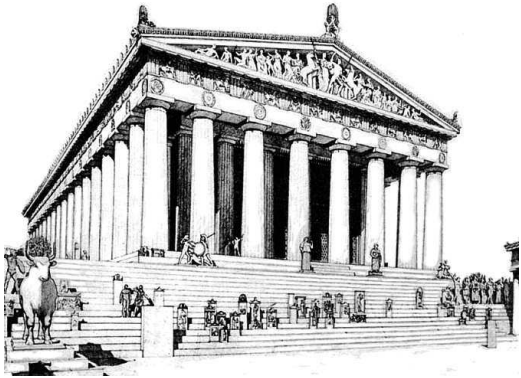}
  \caption{A reconstruction of the entrance of the Parthenon. There 
is considerable discussion about why Athenian culture encouraged 
philosophy, but a popular theory says that it occurred 
because Athens had a direct democracy of citizens, albeit excluding
women, metics and slaves.}
\end{figure}

\subsection{Why (natural) philosophy was born in ancient Greece?}

The transition from {\em Mythos to Logos}, the development of arts and the
birth of democracy are likely related phenomena in ancient Greek society. 
Science, the successor to natural philosophy in modern societies, has 
inherited many of the moral commitments associated with the beginning of 
rational thought, particularly regarding free-thinking and human rights in 
open societies \cite{rights}.

Several circumstances led together to the birth of
western philosophy in ancient Greece
and not elsewhere: \cite{sanchis}
\begin{itemize}
\item Mild weather and plentiful 
spare time (mainly because of slave labour), 
permitting sky observations and outdoors (public) life. 
\item A privileged geographical situation in the Mediterranean sea, favouring 
the contact 
with other peoples and cultures, e.g. in Egypt and Babylon.
\item Commercial relations. Thinkers like Thales were also men of action, 
with curious and enterprising spirits. In fact, Athens became
a maritime power during the era of Pericles. 
\item Self-confidence, freedom of speech, and ultimately 
democracy, with intense participation
of citizens in public life.   
\item Anthropomorphic religion, without holy books and a sacerdotal caste. 
\item Absence of a central power, in sharp contrast to 
the Persian empire, contributing to the development
of local schools of thought and (self) criticism.
\item A flexible and precise (written) language, 
allowing the expression of abstract concepts, e.g. atom by Democritus.
\end{itemize}

At the same time, other circumstances hindered the evolution
of natural philosophy into modern science, e.g.
\begin{itemize}
\item General neglect of experimentation 
(with few exceptions like Archimedes or Hero
of Alexandria in the Hellenistic period), as 
manual labour was generally associated with slavery.
\item Excessive emphasis by many schools 
on the moderation (of customs) as a route to happiness, 
thereby preventing society from technological improvements
leading to a better quality of life. 
\item Dogmatism of schools and 
diversion of mathematical knowledge towards rather 
mystic goals, like in the Pythagorean school. 
\end{itemize}

Many of the above remarks are in order today, since 
intellectual freedom, (self)criticism and dialogue
between schools of thinking lie 
at the heart of scientific progress.

\begin{figure}
  \includegraphics[height=.23\textheight]{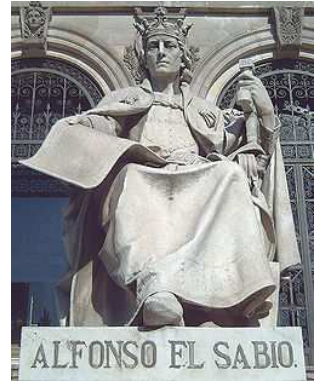}
  \caption{Statue of Alfonso X $\lq\lq$the Wise'' in Madrid.}
\end{figure}

\subsection{From Dark Ages to Modern Science}

According to traditional history, 
during the Dark Ages (denoting here 
the entire period between the fall of Rome and the Renaissance)
scientific knowledge (and culture) 
was almost completely swept away in Western Europe.
However, current scholars tend to avoid 
the term altogether for its negative connotations, 
finding it misleading and inaccurate for some parts of the Middle Ages.

In the Iberian peninsula, the Castilian monarch Alfonso X, 
nicknamed $\lq\lq$the Wise'', 
was one of these exceptions. 
From the beginning of his reign, Alfonso employed Jewish, Muslim and 
Christian scholars at his court. The scientific treatises compiled 
under Alfonso's patronage were the work of the 
celebrated "School of Translators" 
of Toledo. The {\em Alfonsine Tables} are particularly well-known,
containing diagrams and figures on planetary movements.
Because of this, the lunar crater Alphonsus is 
named after him. 

A famous apocryphal quote attributed to Alfonso upon 
hearing an explanation of the complicated assumptions
required in the Ptolemy's view of astronomy was:
\begin{flushright}
{\em If the Lord Almighty had 
consulted me \\ 
before embarking upon creation, I should \\
have recommended something simpler.}
\end{flushright}
Indeed, astronomy was called on to play a crucial role in the birth
of modern science in the following centuries.

\begin{figure}
  \includegraphics[height=.19\textheight]{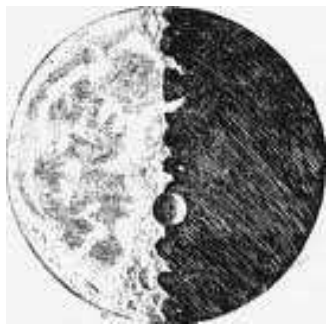}
  \caption{Picture of the Moon from Galileo's $\lq\lq$Starry Messenger''. 
The small book caused sensation on account of the engravings 
it contained showing the Moon's uneven surface.}
\end{figure}

Galileo Galilei published {\em Sidereus Nuncius} (the 
$\lq\lq$Starry Messenger'') in 1610, 
reporting his observations of the Moon and, particularly, his discovery 
of four satellites around Jupiter. 
The lunar observations showed that the surface of the moon was neither 
smooth nor perfectly spherical, but was covered 
with craters and mountains. 
Both discoveries were 
blows to the Aristotelian world-view which was geocentric and maintained 
that everything above the Earth was perfect and incorruptible. The door 
for the return of the old atomistic idea of Democritus 
and Epicurus had been definitely open.  

\subsection{Nuclear beta decay: a compendium of modern physics}

The correct interpretation of X-rays by R\"{o}ntgen together
with the discovery of radioactivity by Becquerel at the end
of the twentieth century was an outstanding step forward in modern 
science. The atom was not to be considered as indivisible, 
in spite of its etymology!

Below we focus on nuclear beta decay as a good example
of the conceptual entanglement of many different issues 
in the scientific 
development. First of all, note that
the {\em ad hoc} hypothesis of the existence of an unseen
particle postulated by Pauli (called neutrino by Fermi), 
to preserve energy-momentum conservation in beta
decay, represents a landmark in the history of physics of the
twentieth century. 

For example,
\begin{itemize}
\item Fermi effective theory brought on the formulation
of a pre-gauge theory with a
V-A structure, which was later interpreted in terms of the
$Z^0$ and $W^{\pm}$ mediators. 
\item The solution to the problem of unitarity violation
in the early theory led ultimately to the discovery of 
non-abelian local gauge invariance (Yang-Mills theories).
\item  The existence of anti-matter postulated by
Dirac is checked in $\beta^+$ decay.
\item More recently, neutrino oscillations
have been the first evidence of physics beyond the SM. 
\item Neutrino physics is called to play a crucial role
in astro-particle physics and cosmology.
\end{itemize}

\begin{figure}
  \includegraphics[height=.19\textheight]{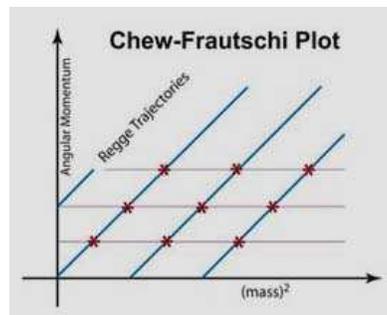}
  \caption{Regge trajectories motivating the string description
of hadrons.}
\end{figure}

\section{From hadronic states to string theory}

Originally, string theory arose in the late 1960s 
in an attempt to understand 
various features of the strong interaction and properties of
hadronic states (see Fig.5). 
The seed of the idea can be traced to the dual resonance
model proposed by Gabriele Veneziano. The picture was that,
for example, a meson should 
consist of two quarks tied together by a piece of `string',
acting as a kind of rubber!

In the early 1970s another theory of the strong interaction, 
Quantum Chromodynamics (QCD), 
was developed, and proved very successful indeed. QCD is
a gauge non-abelian quantum field theory exhibiting 
some peculiar properties depending on the
energy scale: confinement (still to be proved)
and asymptotic freedom. 

Effective theories, like NRQCD \cite{Bodwin:1994jh} 
or HQET \cite{Neubert:1993mb}, 
have become a fruitful 
approach for different kinematic regimes and mass limits. Let us also mention
lattice QCD \cite{lattice} 
as a well-established non-perturbative approach 
formulated on a grid or lattice of points in space and time.

As a result of all this, 
as well as various technical problems with the primitive string 
theory approach, string theory fell out of fashion.
In the last few decades, however,  
string theory \cite{greene,string}
has turned out to be well suited for an even 
more ambitious purpose: to become the $\lq\lq$theory of everything''! 
The basic idea is that
elementary particles or force carriers
should not be described as point-like, but instead
viewed as different oscillation modes of a string - 
the ultimate constituent of nature. 

Think of a guitar string. Depending on how the string is plucked 
and how much tension is in the string, different musical notes will be 
created. In a similar manner, the elementary 
particles could be thought of as the 
excitation modes of elementary strings. 
Furthermore, the spectrum of
string excitations includes a massless particle with two units
of spin which can be identified with the graviton, the expected
carrier of the quantum gravitational field. Therefore, string theory 
could unify all four known forces of Nature, including gravitation,
the unfulfilled dream of many generations of physicists.

Moreover, higher dimensional 
objects (branes) were included in the framework
by Joseph Polchinski \cite{Polchinski}, greatly increasing the 
number of possible background geometries and leading to enormous
consequences in later developments of string theory.

If string theory is to be a theory of quantum gravity, then the average 
size of a string should be somewhere near the Planck length 
($\sim 10^{-35}$ m). 
This means that strings would be, in principle, too small to 
be directly detected by current or expected particle physics experiments. 
Cleverer methods of testing the 
theory than just looking for little strings 
have to be devised in particle experiments. Nonetheless, other 
possibilities have to be considered, including TeV stringy effects
to be potentially detected at the Large Hadron Collider (LHC).

On the other hand, supersymmetry is required 
in order to include fermions in string theory (stabilizing the model too). 
Supersymmetric partners to currently known 
particles are supposed to be too massive for detection
at accelerators till now. However, 
the LHC could be 
on the verge of finding evidence for supersymmetry.

An intriguing feature of string theory 
is that more than three spacial dimensions exist
for self-consistency of the theory. This is not really a new idea
in physics, with Kaluza's early
work showing that general relativity in five dimensions yields 
electromagnetism. More generally, string theory 
provides a theoretical arena in which different compactifications
of extra dimensions lead to many different possibilities for 
physics beyond the SM. Note that extra dimensions
can be at reach of the LHC experiments looking for
a missing energy signal \cite{LHCextra}.

Despite all the beautiful and suggestive features 
(e.g. dualities) of string theory, 
there is so far one main problem with it: string theorists end up with
a landscape of many possible consistent 
solutions ($10^{500}$ or so), each one with a different vacuum
and physical predictions.
In all frankness, this should be somewhat discouraging 
for string theorists \footnote{Leaving aside the 
anthropic principle as an arguable way out \cite{Susskind:2003kw}.}.

\begin{figure}
  \includegraphics[height=.32\textheight]{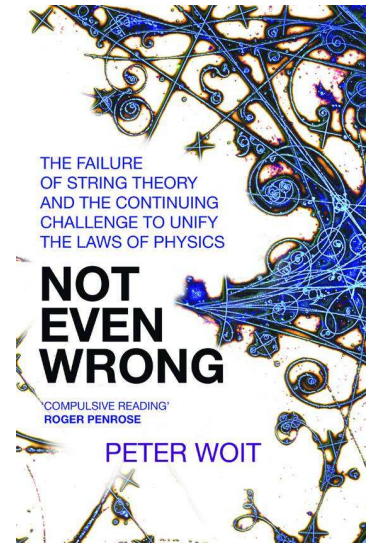}
  \caption{The epistemological difficulties of the vacua 
landscape in string theory
has led many physicists to become rather skeptical about it. 
The title of the book by Peter Woit \cite{Woit} 
{\em Not even wrong}, corresponds to the  
Pauli's term for scientifically useless speculative theories
(i.e. cannot be falsified).}
\end{figure}

\subsection{Strings strike back: Maldacena's holographic principle}

Gerard 't Hooft's 
visionary proposal of the holographic principle \cite{Stephens:1993an}
was in part an inspiration for Juan Maldacena's conjecture. Basically, 
everything that happens in a given volume of space can be
represented as taking place (is encoded) 
on a surface surrounding this volume \cite{Susskind:1994vu}.

Maldacena's conjecture \cite{Maldacena:1997re,Gubser:1998bc,Witten:1998qj}
proposes that a string theory
on a certain background spacetime
is equivalent to a conformal supersymmetric
Yang-Mills theory on a lower dimension (AdS/CFT correspondence).
Such a duality implies:
\begin{flushright}
Strongly coupled dynamics $\Leftrightarrow$ Weakly coupled strings
\end{flushright}
and vice versa.

In particular, one may think of QCD as a scale-invariant theory
(QCD is almost conformal at momenta much larger than $\Lambda_{QCD}$)
with a useful correspondence with a certain string theory on a
anti-de Sitter spacetime and a closed manifold (like a five dimensional 
sphere). 

Can the dual string theory explain confinement, hadronization, etc?

It should be emphasized that the gauge/string duality has 
not yet been proved but many non-trivial tests have been carried out.
Some physicists do believe that holography is a fundamental
property of string theory.
Finally notice that recent data from 
heavy ion processes at RHIC might be interpreted 
by invoking a dual string theory \cite{Horowitz:2010dm}.

\begin{figure}
  \includegraphics[height=.17\textheight]{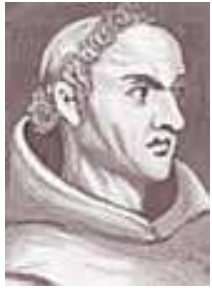}
  \caption{Ockham's razor 
({\em entia non sunt multiplicanda praeter necessitatem})  
 is popularly expressed as 
$\lq\lq$the simplest explanation is more likely the correct one''.
The exact meaning of $\lq\lq$simplest'' must be nuanced however.}
\end{figure}

\section{New symmetries, new worlds}

Besides the string description, the SM can obviously 
be extended in many different ways.  
There are reasonable and motivated scenarios,
most of them including supersymmetry since it provides
a solution for the Higgs mass problem \cite{Drees:1996ca}. 
Alternative popular
scenarios are, e.g., the little Higgs model \cite{Schmaltz:2005ky},  
warped extra dimensions \cite{Randall:1999ee}
or technicolor \cite{Farhi:1980xs}. 

Minimal extensions of the SM can be seen as 
a theoretical prejudice attached to beauty and elegance
of the formalism, also supported so far by many successes 
in all fields of science. Indeed, one should not forget 
the scientific virtues usually attributed to
a healthy economy of hypotheses ({\em Ockham's razor}). 

However, let us point out as a caveat
that the third generation of quarks and leptons, 
in the early SM, was in fact not demanded by previous 
experimental data nor theoretical requirements. 
In this regard, Einstein warned: 
\begin{flushright}
{\em Everything should be as simple as possible,\\ 
but not simpler.}
\end{flushright}

Among those possible not so much strongly motivated scenarios
beyond our present conventional wisdom, new kinds of 
strong interaction have been considered 
in particle physics \cite{Okun:1980mu,Kang:2008ea}. 
Actually, the existence of additional gauge groups with matter
in the fundamental representations would arise naturally in
string theory \cite{ArkaniHamed:2005yv}. 

For instance, the usual $SU(3)\times SU( 2)\times U(1)$ group 
of the SM can be extended by a new non-abelian group 
(like in hidden valley models \cite{Strassler:2006im}). 
Higher dimension operators at the TeV scale
(induced by a $Z'$ or by a loop of heavy particles
carrying charges from both the SM and the new group)
should allow interactions between the SM fields and the
new particles. 

Rather exotic phenomenology would 
show up at the LHC: high multiplicity events, displaced
vertices and missing energy, long range correlations, etc  
\cite{Han:2007ae,SanchisLozano:2008te}.

\section{From the $\lq\lq$small'' to the $\lq\lq$large'' 
and vice versa}

Astronomical observations played a leading role 
in the early development of natural philosophy in ancient Greece
(as in previous civilizations) and
the scientific revolution by Galileo and Newton 
(alongside many others). Certainly, $\lq\lq$messengers from the
sky'' (light, cosmic rays, dark matter ...)
will still provide us with essential information from/on
the Cosmos.

Fundamental questions in physics unite the largest
scale (the study of the universe) to the smallest scale (elementary
particle physics). In particular, the nature of dark matter 
\cite{Bertone:2004pz}
directly involves both cosmology and certain extensions of the
SM (like supersymmetric models) 
which furnish candidates (e.g. neutralino, gravitino) for the so-called 
Weak Interacting Massive Particle (WIMP).

\begin{figure}
  \includegraphics[height=.27\textheight]{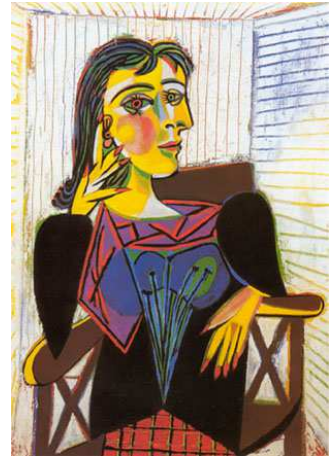}
  \caption{Portrait of Dora Maar by Picasso. 
New symmetries (and their breaking) should
furnish a guide useful in the search for physics beyond
the SM.}
\end{figure}

If the Milky Way's dark matter halo is composed of WIMPs,
the WIMP flux on Earth should be about $10^5$ cm$^{-2}$s$^{-1}$.
Therefore direct detection experiments should be able to detect
WIMP elastic scattering off nuclei, providing a signal
if backgrounds are low enough.

The WIMP-nucleon (nucleus) cross section (spin-dependent and spin-independent)
encodes:
\begin{enumerate}
\item
  Particle physics inputs, including WIMP interaction properties
\item
  Nuclear properties: 
hadronic matrix elements describing the quark and 
gluon content of the nucleon, notably concerning strange quarks.
\item
  Astrophysics: 
   WIMP velocity distribution, Earth motion.
\end{enumerate}
This is an illustrative example of the multiple fields involved in
any claim of observation of dark matter.

\section{Concluding Remarks}

As history teaches us, it is hardly conceivable a steady development of 
science which leaves aside any area of knowledge. In this paper, 
we have illustrated this point of view with 
several examples from both elementary particle and astro-particle physics,
as well as cosmology. 

Returning to Hilbert's quotation, which
entitles this paper, one might complete it by saying:
\begin{flushright}
{\em We must know $\dots$ about everything.}
\end{flushright} 

In answering these pending big questions about Nature, we 
should (hopefully)
become more $\lq\lq$human'', that is, 
wiser in the broad sense of the word (committed in world-wide solidarity,  
respectful towards the environment...). 
After all, we should live up to 
the name of our species: {\em Homo sapiens}, $\lq\lq$ knowing man''.


\begin{theacknowledgments}
  I would like to thank the organizers of the IX Conference 
on Quark Confinement and the Hadron Spectrum in Madrid
for inviting me
to give the after-dinner talk and write this somehow non-standard paper for
the Proceedings. I acknowledge support from MICINN (FPA2008-02878)
and Generalitat Valenciana (GVPROMETEO2010-056). 
I dedicate the after-dinner talk and this
paper to the memory of Francisco Yndurain and Ximo Prades.  
\end{theacknowledgments}



\bibliographystyle{aipproc}   

\bibliography{sample}


\end{document}